# One-pot green process to synthesize controllable surface terminations MXenes in molten salts

Miao Shen[a, b, c], Weiyan Jiang, Liang Guo, Sufang Zhao, Rui Tang, and Jianqiang Wang[a, b, c,]

a Shanghai Institute of Applied Physics, Chinese Academy of Sciences, Shanghai 201800, China
b Dalian National Laboratory for Clean Energy, Dalian 116023, China
c Key Laboratory of Interfacial Physics and Technology, Chinese Academy of Science, Shanghai 201800, China

Surface terminations for 2D MXene have dramatic impacts on physicochemical properties. The commonly etching methods usually introduce -F surface termination or metallic into MXene. Here, we present a new molten salt assisted electrochemical etching (MS-E-etching) method to synthesize fluorine-free $Ti_3C_2T_x$ without metallics. Due to performing electrons as reaction agent, the cathode reduction and anode etching can be spatially isolated, thus no metallic presents in $Ti_3C_2T_x$ product. Moreover, the $T_x$ surface terminations can be directly modified from -Cl to -O and/or -S in one pot process. The obtained -O terminated MXenes exhibited capacitance of 225 and 205 F/g at 1 and 10 A/g, confirming high reversibility of redox reactions. This one-pot process greatly shortens the modification procedures as well as enriches the surface functional terminations. More importantly, the recovered salt after synthesis can be recycled and reused, which brands it as a green sustainable method.

## Introduction

Due to environmental issues and the consumption of fossil fuels, the renewable and sustainable energy is devloping rapidly. Accordingly, high efficiency energy conversion and storage (ECS) technology, has attracted wide attention in recent decades [1-2]. Among which, a new type of two-dimensional (2D) transition metal carbides or nitrides, called MXenes are regarded as excellent candidates for ECS due to their unique physicochemical properties [3-6]. MXenes are written as $M_{n+1}X_nT_x$ (n=1-3) since they are usually prepared by selective etching of A layer elements (ⅢA and ⅣA group elements) in MAX phase precursors, where M represents an early transition metal element, X is carbon and/or nitrogen, $T_x$ stands for surface terminations.

Unlike the surfaces of other 2D materials, such as graphene and transition metal dichalcogenides, the $T_x$ for MXene can be chemically modified, which will have dramatic impacts on material stabilities, electronic structures and other physicochemical properties [7-10]. The commonly etching methods to synthesis MXene usually introduce functional group of -F [3, 5, 11], which has harmful influence on conductivity and damages its applications in ECS [1, 12]. Lewis acid salt replacement method has been recently applied for etching MAX phases without -F terminate, however, the reduced metallic is inevitably presented in MXene product [13, 14]. For further application or surface modification, the metallic has to be removed by washing with oxidizing acid [15], followed by post-thermal annealing or solvothermal treated [16-23], which complicates the operations and produces waste liquid as well. Therefore, it is highly desirable to develop a simple and environmentally sound process to prepare MXene with controllable functional groups.

Herein, we present the first work on a new molten salt assisted electrochemical etching (MS-E-etching) method to synthesize fluorine-free $Ti_3C_2T_x$. Due to performing electrons as reaction agent, the cathode reduction and anode etching can be spatially isolated, thus no metallic presents in $Ti_3C_2T_x$. Furthermore, the $T_x$ surface terminations can be directly modified in this molten salt system, without producing any waste liquid in the process. The one pot preparation method greatly shortens the modification procedures and enriches the surface functional terminations. More importantly, the recovered salt after synthesis can be recycled and reused, which brands it as a sustainable method [24-26].

## Results and Discussion

## Synthesis of $Ti_3C_2T_x$ by MS-E-etching

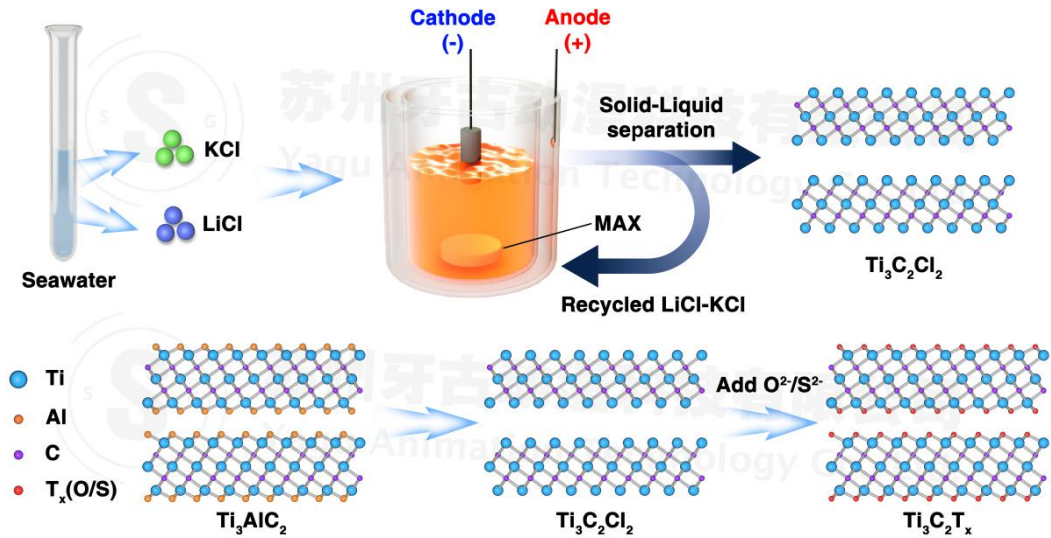

Fig.1 Synthesis sketch of MXene from MAX phase by MS-E-etching and the subsequent controllable surface terminations

Due to abundant lithium and potassium resources in salt lakes, eutectic salts of LiCl and KCl have a low melting point, we propose MS-E-etching $Ti_3AlC_2$ in LiCl-KCl (1:1 wt%) salts at 450℃ (Fig.1). The synthetic details are summarized in the Supporting information. When a potential of 2.0 V is applied, the $Ti_3AlC_2$ phase releases electrons to form some active sites, and -Cl absorbs onto these active sites bonding with Al, due to their strong binding capability [27]. $AlCl_3$ has a boiling point of 180 °C and is expected to rapidly evaporate at 450 °C, which provides the driving force for the outward diffusion of the Al atom. Meanwhile, some of the $Cl^-$ anions spontaneously bond with $Ti_3C_2$ to form a more stable phase (i.e., $Ti_3C_2Cl_2$) and result in the release of extra electrons. Simultaneously, metallic Li or K deposited on cathode (**Supporting information, Fig. S1**), the overall MS-E-etching reaction can be described as follows:

Anode: $Ti_3AlC_2 + 3Cl^- = AlCl_3\uparrow + Ti_3C_2 + 3e^-$　　　　　　　　　　(1)

　　　　$Ti_3C_2 + xCl^- = Ti_3C_2Cl_x + xe^-$　　　　　　　　　　　　　(2)

Cathode: $Li^+(K^+) + e^- = Li(K)$　　　　　　　　　　　　　　　　(3)

Overall reaction: $Ti_3AlC_2 + (3+x) LiCl (KCl) = Ti_3C_2Cl_x + (3+x) Li (K) + AlCl_3\uparrow$　　(4)

Reaction (1) is essential to exfoliate Al layers from $Ti_3AlC_2$ anode. Due to the evaporation of $AlCl_3$, this process at high temperature is superior to etching in aqueous solution that usually occurs preferentially on the surface [28]. The overall etching yield, calculated by the weight ratio of etched material to the bulk precursor, is around 94.6 % (**Table S1 and Fig. S2**).

## Characterization of MS-E-etching $Ti_3C_2T_x$

X-ray diffraction (XRD) patterns of the pristine $Ti_3AlC_2$ before (black), and after MS-E-etching at 450 °C for 24 h (red) are shown in Fig. 2a. The characteristic peak (002) of $Ti_3AlC_2$ at $2\theta=9.8º$ shifted to a lower angle ($2\theta=8.1º$), attributed to an increased c lattice parameter from 18.2 to 21.9 Å; implying the successful etch of $Ti_3AlC_2$ into layered $Ti_3C_2T_x$ (MXene). The obtained $Ti_3C_2T_x$ powders show black colors, which can well disperse in anhydrous alcohol (0.17 mg/mL) and present the clear Tydall effect (Fig. 2a, Insert). A scanning electron microscopy (SEM) image of the MS-E-etching $Ti_3C_2T_x$ sample is shown in Fig. 2b. Multi-layered $Ti_3C_2T_x$ powders remain tightly stacked. The lack of apparent expansion may be ascribed to mild etching condition without violent gas releasing [29]. The corresponding EDS analysis (**Fig. S3**) indicates that the elemental composition is Ti/Cl=1.85 in atomic ratio, with small amounts of Al (1.2 atom %), and O (7.1 atom %). The presence of oxygen seems to come from adsorbed water in the top surface region of sample [14]. The lamellar microstructure of the $Ti_3C_2T_x$ is clearly visible in scanning transmission electron microscopy (STEM) images, each sheet exhibits bright atomic columns in the middle and two dark surface layers (Fig. 2c). Fourier reconstruction and filtering of the labelled part of the micrograph with point symmetry constrained to p2 further emphasize the superstructure (Fig. 2d and Fig. 2e) [30], it shows the same atom arrangement with the simulated $Ti_3C_2Cl_2$ structure (Fig. 2f). NMR results show a resonance at 566 ppm for $Ti_3AlC_2$, which is shifted to 350 ppm for the MS-E-etching $Ti_3C_2T_x$ (Fig. 2g). This shift is most likely due to the difference in surface functionalization -Cl [13]. The X-ray photoelectron spectroscopy (XPS) shows four major bands, Ti 2p, Cl 2p, C 1s, and O 1s, and the Al peak is un-observable after MS-E-etching (**Fig. S4**). High-resolution Ti 2p spectrum (Fig. 2h) shows the peaks at 455.4 and 455.9 eV are assigned to the Ti−C (I) ($2p_{3/2}$) and Ti−C (II) ($2p_{3/2}$) bond [31, 32]. The peaks at 457.4 and 459.8 eV, can be assigned to the Ti−Cl ($2p_{3/2}$) and Ti−O ($2p_{3/2}$) bonds with higher valence [33-34]. The Ti:Cl ratio determined by the XPS analysis is 2.94:2. These results further confirm the formation of $Ti_3C_2Cl_2$ after MS-E-etching in LiCl-KCl melts.

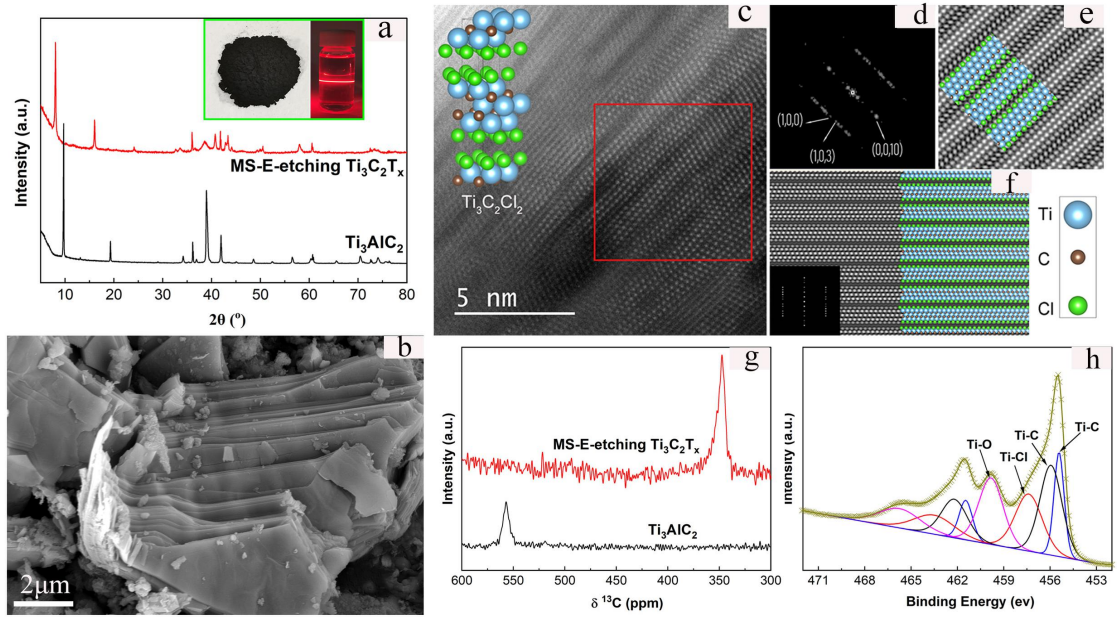

Fig.2 (a) XRD patterns of $Ti_3AlC_2$ and the obtained $Ti_3C_2T_x$, Insert: Optical image of the as-received MS-E-etching powders and the anhydrous alcohol dispersion of $Ti_3C_2T_x$; (b) Representative SEM images of $Ti_3C_2T_x$; (c) HR-STEM image showing the atomic positions of $Ti_3C_2Cl_2$; (d) and (e) Fourier reconstruction and filtering of the labelled part of the micrograph with point symmetry constrained to p2 further emphasize the structure; (f) the simulated atomic arrangement of $Ti_3C_2Cl_2$ structure; (g) NMR images of $Ti_3AlC_2$ and $Ti_3C_2T_x$; (f) High-resolution Ti 2p spectrum of $Ti_3C_2T_x$ powder;

## MS-E-Etching Mechanism by Density-functional theory (DFT)

The etching mechanism of $Ti_3AlC_2$ and its corresponding etching potentials in molten LiCl-KCl were calculated by the classical electrochemical dynamoelectric method [35-37]. The energy bands for $Ti_3AlC_2$ as well as its low density surface (100) are contributed by Ti's d-orbita (**Fig. S5**). After applying electricity, electronic transition from Ti produces active sites, Cl anions in molten salt will attach to these active sites, step by step in the form of $*TiCl_x$ (x=1-3, shown in Eq. 5 to Eq. 7) [38]. If the fourth Cl anion can attach on $*TiCl_3$, stable $*TiCl_4$ (Eq. 8) would be desorbed due to its strong volatility (Eq.9). However, this step only occurs at the applied electrolytic potential higher than 0.91V [37, 38] (**Fig. 3, Fig. S6**). Since Al layer is successfully etched from $Ti_3AlC_2$, it is deduced that the -Cl on $*TiCl_3$ sites can convert to Al sites in the form of $*TiCl_2AlCl$ at high temperature (Eq. 10). Meanwhile, -Cl anion can also attach on $*TiCl_2AlCl$ under the electric field, in the form of $*TiCl_3AlCl$ (Eq. 11), which will subsequently transform to $*TiCl_2AlCl_2$ (Eq. 12). Through stepwise electrochemical adsorption and thermo-chemical transformation, the $*TiCl_2AlCl_3$ eventually forms (Eq. 13 and Eq. 14); and the $AlCl_3$ will be detached and volatilized

(Eq. 15). Hence, the Ti$_3$C$_2$Cl$_2$ is obtained under the coupling effect of electrochemical and thermo-chemical, through controlling the applied potentials at high temperatures.

When the applied potential is lower to 1.6 V, the actual working bias on the anode was 0.002 V (vs. Ag/AgCl) (**Fig. S7**), the product is consisted of Ti$_3$C$_2$Cl$_2$ and Ti$_3$AlC$_2$ raw material (**Fig. S8**), demonstrating that etching of Al occurs as anodic electricity is given. However, as the applied voltage increases to 3.0 V, the actual working bias on the anode was 1.2 V (vs. Ag/AgCl) (**Fig. S7**), only TiC is obtained (**Fig. S8**), which further indicates that some of the titanium will be etched at potential over 0.9 V vs. Ag/AgCl.

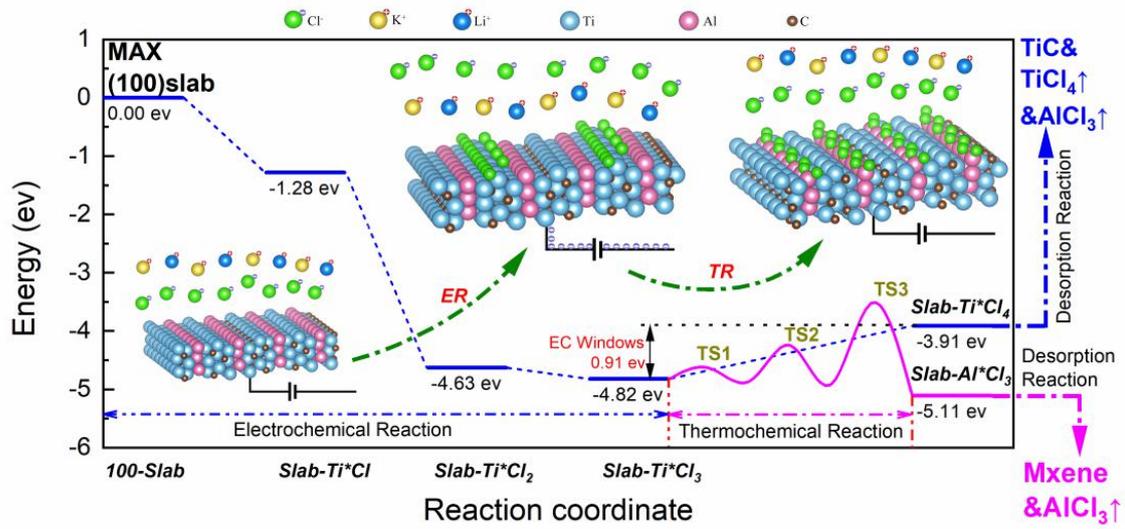

Fig.3 The reaction coordinate of MS-E-Etching Ti$_3$C$_2$T$_x$

$$Slab + Cl^- \rightarrow *TiCl + eU \tag{5}$$

$$*TiCl + Cl^- \rightarrow *TiCl_2 + eU \tag{6}$$

$$*TiCl_2 + Cl^- \rightarrow *TiCl_3 + eU \tag{7}$$

$$*TiCl_3 + Cl^- \rightarrow *TiCl_4 + eU \tag{8}$$

$$*TiCl_4 \rightarrow TiCl_4 \uparrow \tag{9}$$

$$*TiCl_3 \rightarrow *TiCl_2AlCl \tag{10}$$

$$*TiCl_2AlCl + Cl^- \rightarrow *TiCl_3AlCl + eU \tag{11}$$

$$*TiCl_3AlCl \rightarrow *TiCl_2AlCl_2 \tag{12}$$

$$*TiCl_2AlCl_2 + Cl^- \rightarrow *TiCl_3AlCl_2 + eU \tag{13}$$

$$*TiCl_3AlCl_2 \rightarrow *TiCl_2AlCl_3 \tag{14}$$

$$* TiCl_2AlCl_3 \rightarrow * TiCl_2 + AlCl_3 \uparrow \qquad (15)$$

## Substitution of Cl surface terminations

After MS-E-etching, the cathode was withdrawn from electrolytic cell. Another type of inorganic salt, such as Li$_2$O and Li$_2$S was pressed into pellet and directly added into this melting system to modify the -Cl surface terminations for other atoms (**Fig. 1**). The characteristic peak (002) of MS-E-etching Ti$_3$C$_2$Cl$_2$ further shifted from 8.1º to a lower angle 7.4º (Li$_2$O treated) and 6.8º (Li$_2$S treated), attributed to an increased c lattice parameter from 21.9 Å to 23.6 and 25.9 Å (**Fig. 4a**). High-resolution Ti 2p spectrum (**Fig. 4b**) also demonstrates the absence of Ti−Cl bonds after Ti$_3$C$_2$Cl$_2$ treated in LiCl-KCl-Li$_2$O, but a new peak at 458.9 eV is detected, which can be assigned to the Ti−O (2p$_{3/2}$) bond. For Li$_2$S treated Ti$_3$C$_2$T$_x$, the Ti−Cl bond is also not observed, two peaks at 459.2 eV and 456.8 eV appear, which can be assigned to Ti-O (2p$_{3/2}$) and Ti-S bonds (2p$_{3/2}$) [15]. These results indicate that the -Cl surface termination is transformed to -O and/or -S functional groups. The corresponding SEM confirms the modified Ti$_3$C$_2$T$_x$ powders still remain multi-layered and tightly stacked (**Fig. 4c, Fig. 4d**). EDS analysis further demonstrates the element Cl is replaced by element O or/and S (**Fig. S9, Fig. S10**). Moreover, we attempted to explore the MS-E-etching process to synthesize MXene from MAX with other A site layer element. Ti$_3$C$_2$T$_x$ was successful synthesized from Ti$_3$SiC$_2$, of which the characterizations were provided in the Supporting Information (**Fig. S11**).

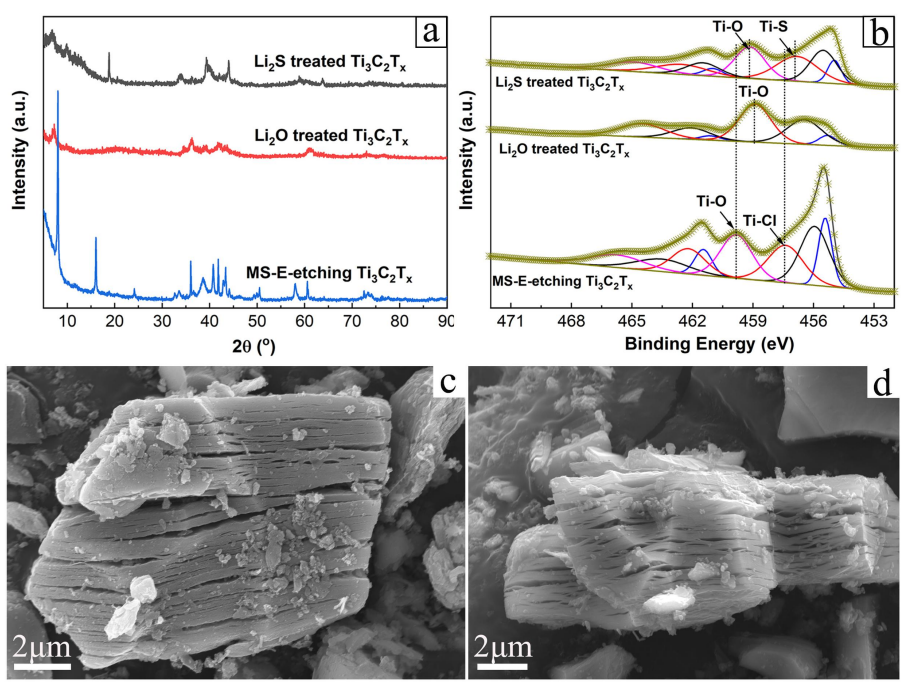

Fig. 4 (a) XRD patterns of obtained $Ti_3C_2T_x$ with various surface terminations; (b) High-resolution Ti 2p spectrum of MS-E-etching$Ti_3C_2T_x$ after treated in LiCl-KCl-$Li_2O$ and LiCl-KCl-$Li_2S$; (c) Representative SEM images of $Ti_3C_2T_x$ after treated in LiCl-KCl-$Li_2O$; (d) SEM images of $Ti_3C_2T_x$ after treated in LiCl-KCl-$Li_2S$

## Electrochemical Properties

It is well known that $Ti_3C_2T_x$ is an excellent material for supercapacitors due to its pseudocapacitive storage mechanism [8], and surface terminations are crucial for the electrochemical performance [39-43]. Here, the cyclic voltammetry (CV) and galvanostatic charge-discharge (GCD) experiments were conducted in the three-electrode system to investigate the electrochemical properties of $Ti_3C_2T_x$ with different surface functional groups. As shown Fig. 5a, with the same mass loading and scan rate, the CV curves show different areas and different levels of stored charges, the capacitance for $Li_2O$ treated $Ti_3C_2T_x$ distincitively increased. Fig. 5b presents the GCD curves of $Li_2O$ treated $Ti_3C_2T_x$ at different current densities, showing the capacitance corresponds to 223 F/g and 205 F/g, at 1A/g and 10A/g, respectively, confirming the high reversibility of redox reactions at different current densities (**Fig. 5c**). Such capacitance is comparable to many modified composite materials (**Table S2**). Moreover, the $Li_2O$ treated $Ti_3C_2T_x$ electrode withstood 10000 cycles with the retention above 100% at 10 A/g, showing that the $Ti_3C_2T_x$ was gradually delaminated during the cycle test and more ions intercalation and deintercalation.

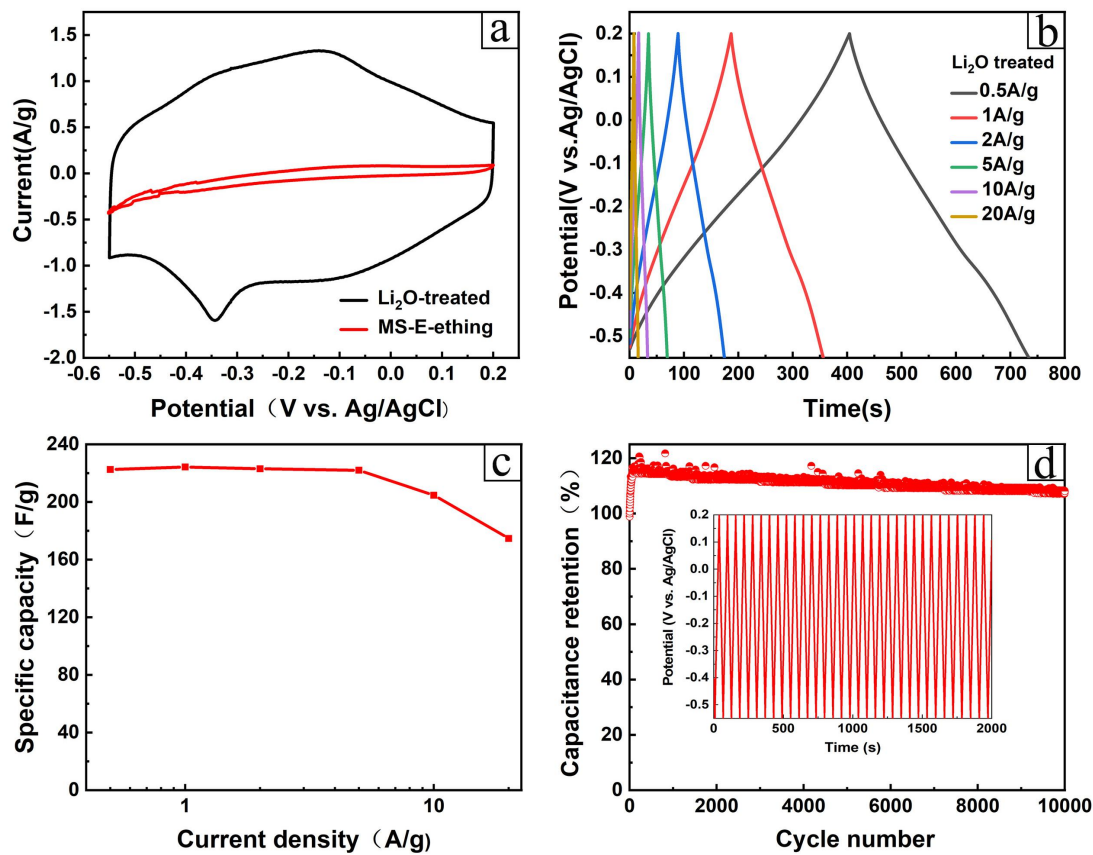

Fig.5 Electrochemical performance of $Ti_3C_2T_x$ electrode in three-electrode system. a) Cyclic voltammetry for MS-E-etching $Ti_3C_2T_x$ and $Li_2O$ treated $Ti_3C_2T_x$ at 5mV/s in 1M $H_2SO_4$; b) The galvanostatic charge/discharge curves for $Li_2O$ treated $Ti_3C_2T_x$ at different current densities in 1M $H_2SO_4$; c) The specific capacitance of $Li_2O$ treated $Ti_3C_2T_x$ at different current densities; d) Capacitance retention of $Li_2O$ treated $Ti_3C_2T_x$ electrode in 1M $H_2SO_4$. Insert shows galvanostatic cycling data collected at 10A/g.

In summary, we present a new molten salt assisted electrochemical etching (MS-E-etching) method to synthesize $Ti_3C_2T_x$ without metallics. By controlling the electrolysis voltage lower than 0.91V vs. Ag/AgCl in LiCl-KCl salts at high temperatures, the selective removal of Al atoms proceeds with producing the ordered crystal $Ti_3C_2Cl_2$. After withdrawing the cathode from electrolytic cell and addition of various inorganic salt, the surface terminations can be directly modified from -Cl to -O and/or -S in this one pot molten salts without producing any waste liquid. The obtained -O terminated MXene exhibits capacitance of 225 and 205 F/g at 1 and 10 A/g, confirming high reversibility of redox reactions. This one-pot process greatly shortens the modification procedures as well as enriches the surface functional terminations. And more importantly, the recovered salt after synthesis can be recycled and reused, which brands it as a green sustainable method.

## Acknowlegements

This work was supported by Natural Science Foundation of Shanghai (Grant No.19ZR1468300); and the "Transformational Technologies for Clean Energy and Demonstration", Strategic Priority Research Program of the Chinese Academy of Sciences (Grant No. XDA 21000000).

# References

[1] B. Anasori, M. R. Lukatskaya, Y. Gogotsi, *Nature Rev. Mater.* **2017**, 2, 16098.

[2] V. Nicolosi, M. Chowalla, M. G. Kanatzidis, M. S. Strano, J. N. Coleman, *Science.* **2013**, 340, 1226419.

[3] M. Naguib, M. Kurtoglu, V. Presser, J. Lu, J. Niu, M. Heon, L. Hultman, Y. Gogotsi, M. W. Barsoum, *Adv. Mater.* **2011**, 23, 4248-4253.

[4] M. Naguib, V. N. Mochalin, M. W. Barsoum, Y. Gogotsi, *Adv. Mater.* **2014**, 26, 992-1005.

[5] M. Ghidiu, M. R. Lukatskaya, M. Q. Zhao, Y. Gogotsi, M. W. Barsoum, *Nature.* **2014**, 516 (7529), 78−81.

[6] F. Shahzad, M. Alhabeb, C. B. Hatter, B. Anasori, S. M. Hong, C. M. Koo, Y. Gogotsi, *Science.* **2016**, 353 (6304), 1137−1140.

[7] Y. Liu, H. Xiao, W. A. Goddard, *J. Am. Chem. Soc.* **2016**, 138, 15853–15856 (2016).

[8] C. Si, J. Zhou, Z. Sun, *ACS Appl. Mater. Interfaces.* **2015**, 7, 17510–17515.

[9] M. Khazaei, M. Arai, T. Sasaki, C.-Y. Chung, N. S. Venkataramanan, M. Estili, Y. Sakka, Y. Kawazoe, *Adv. Funct. Mater.* **2013**, 23, 2185–2192.

[10] L. Zhou, Y. Zhang, Z. Zhuo, A. J. Neukirch, S. Tretiak, *J. Phys. Chem. Lett.* **2018**, 9, 6915–6920.

[11] J. Halim, M. R. Lukatskaya, K. M. Cook. *Chem.Mater.* **2014**, 26, 2374-2381.

[12] M. A. Hope, A. C. Forse, K. J. Grififith, *Phys. Chem. Chem. Phys.* **2016**, 18, 5099-5102.

[13] M. Li, J. Lu, K. Luo, Y. Li, K. Chang, K. Chen, J. Zhou, J. Rosen, L. Hultman, P. Eklund, P. O. Å. Persson, S. Du, Z. Chai, Z. Huang, Q. Huang, *J. Am. Chem. Soc.* **2019**, 141, 4730–4737.

[14] Y. Li, H. Shao, Z. Lin, J. Lu, L. Liu, B. Duployer, P. O. Å. Persson, P. Eklund, L. Hultman, M. Li, K. Chen, X. H. Zha, S. Du, P. Rozier, Z. Chai, E. Raymundo-Piñero, P. L. Taberna, P. Simon, Q. Huang, *Nature Mater.* **2020**, 19, 894–899.

[15] V. Kamysbayev, A. S. Filatov, H. Hu, X. Rui, F. Lagunas, D. Wang, R. F. Klie, D. V. Talapin. *Science,* **2020**, 369, 979-983.

[16] R. B. Rakhi, B. Ahmed, M. N. Hedhili, D. H. Anjum, H. N. Alshareef, *Chem. Mater.* **2015**, 27, 5314-5323.

[17] S. Lai, J. Jeon, S. K. Jang, J. Xu, Y. J. Choi, J.-H. Park, E. Hwang, S. Lee, *Nanoscale.* **2015**, 7, 19390-19396.

[18] K. Wang, Y. Zhou, W. Xu, D. Huang, Z. Wang, M. Hong, *Ceram. Int.* **2016**, 42, 8419-8424.

[19] M. Han, X. Yin, H. Wu, Z. Hou, C. Song, X. Li, L. Zhang, L. Cheng, *ACS Appl. Mater. Interfaces.* **2016**, 8, 21011;

[20] X. T. Gao, Y. Xie, X. D. Zhu, K. N. Sun, X. M. Xie, Y. T. Liu, J. Y. Yu, B. Ding, *Small.* **2018**, 14, 1802443;

[21] H. Pan, X. Huang, R. Zhang, D. Wang, Y. Chen, X. Duan, G. Wen, *Chem. Eng. J.* **2019**, *358*, 1253;

[22] S. Lai, J. Jeon, S. K. Jang, J. Xu, Y. J. Choi, J.-H. Park, E. Hwang, S. Lee, *Nanoscale.* **2015**, *7*, 19390.

[23] R. B. Rakhi, B. Ahmed, M. N. Hedhili, D. H. Anjum, H. N. Alshareef, *Chem. Mater.* **2015**, *27*, 5314;

[24] X. Xiao, H. B. Song, S. Z. Lin, Y. Zhou, X. J. Zhan, Z. M. Hu, Q. Zhang, J. Y. Sun, B. Yang, T. Q. Li, L. Y. Jiao, J. Zhou, J. Tang, Y. Gogotsi, *Nat. Commun.* **2016**, 7, 11296.

[25] W. Weng, B. Jiang, Z. Wang, W. Xiao, *Sci. Adv.* **2020**, 6, eaay9278.

[26] X. Liang, J. Xiao, W. Weng, W. Xiao, *Angew. Chem. Int. Ed.* **2020**, DOI: 10.1002/anie.202013257;

[27] R. S. Alwitt, H. Uchi, T. R. Beck, R. C. Alkire, *J. Electrochem. Soc.* **1984**, 131, 13-17.

[28] W. Sun, S. A. Shah, Y. Chen, Z. Tan, H. Gao, T. Habib, M. Radovic, M. J. Green. *J. Mater. Chem. A*. **2017**, 5, 21663

[29] M. Alhabeb, K. Maleski, B. Anasori, P. Lelyukh, L. Clark, S. Sin, Y. Gogotsi. *Chem. Mater.* **2017**, 29, 7633−7644

[30] J. Xu, T. White, P. Li, C. He, J. Yu, W. Yuan, Y. Han, *J. Am Chem. Soc.* **2010**, 132, 10398-10406

[31] J. Halim, K. M. Cook, M. Naguib, P. Eklund, Y. Gogotsi, J. Rosen, M. W. Barsoum, *Appl. Surf. Sci.* **2016**, 362, 406-417.

[32] V. Schier, J. Halbritter, K. Karlsruhe, ARXPS-analysis of sputtered TiC, SiC and $Ti_{0.5}Si_{0.5}C$ layers. *Fresenius' J. Anal. Chem.* **1993**, 346, 227-232.

[33] E. Magni, G. A. Somorjai, *Surf. Sci.* **1996**, 345 (1-2), 1-16.

[34] C. M. Desbuquoit, J. Riga, J. J. Verbist, *J. Chem. Phys.* **1983**, 79 (1), 26-32.

[35] J. K. Nørskov, J. Rossmeisl, A. Logadottir, L. Lindqvist, J. R. Kitchin, T. Bligaard, and H. Jónsson. *J. Phys. Chem. B*. **2004**, 108, 46, 17886–17892.

[36] J. Rossmeisl, A. Logadottir and J. K. Nørskov. *Chem. Phys.* **2005**, 319, 178-184.

[37] X. Ma, J. Liu, H. Xiao, and J. Li. *J. Am. Chem. Soc.* **2018**, 140, 1, 46–49

[38] J. Liu, H. Xiao, J. Li. *J. Am. Chem. Soc.* **2020**, 142, 3375-3383.

[39] Z. H. Fu, Q. F. Zhang, D. Legut, C. Si, T. C. Germann, T. Lookman, S. Y. Du, J. S. Francisco, R. F. Zhang. *Phys. Rev. B.* **2016**, 94, 104103.

[40] S. Kajiyama, L. Szabova, H. Iinuma, A. Sugahara, K. Gotoh, K. Sodeyama, Y. Tateyama, M. Okubo, A. Yamada, *Adv. Energy Mater.* **2017**, 7 (9), 1601873.

[41] Ho. Yu, Y. Wang, Y. Jing, J. Ma, C. Du, Q. Yan, *Small*. **2019**, *15*, 1901503

[42] M. Hu, H, Zhang, T. Hu, B. Fan, X. Wang, Z. Li, *Chem. Soc. Rev.* **2020**, 49, 6666–6693.

[43] M. K. Aslam, Y. Niu, M. Xu, *Adv. Energy Mater.* **2021**, 11 (2), 2000681.